\documentclass[journal,transmag]{IEEEtran}
\usepackage{cite}

\ifCLASSINFOpdf
  \usepackage[pdftex]{graphicx}
\else
\fi

\usepackage{stfloats}
\fnbelowfloat

\usepackage{fancyhdr}
\usepackage[normalem]{ulem}
\usepackage[hyphens]{url}
\usepackage[final]{microtype}
\usepackage{flushend}
\usepackage{booktabs}
\usepackage{array}
\usepackage{subfig}
\usepackage{tablefootnote}
\usepackage{tabularx}

\usepackage[bookmarks=true,breaklinks=true,letterpaper=true,colorlinks,linkcolor=black,citecolor=blue,urlcolor=black]{hyperref}

\begin{document}
%
\title{Systolic Tensor Array: An Efficient Structured-Sparse\\GEMM Accelerator for Mobile CNN Inference}


\author{\IEEEauthorblockN{Zhi-Gang Liu,
Paul N. Whatmough,
Matthew Mattina}
\IEEEauthorblockA{Arm ML Research Lab, Boston, MA, USA}
}

%



\IEEEtitleabstractindextext{%
\begin{abstract}
Convolutional neural network (CNN) inference on mobile devices demands efficient hardware acceleration of low-precision (INT8) general matrix multiplication (GEMM).
The systolic array (SA) is a pipelined 2D array of processing elements (PEs), with very efficient local data movement, well suited to accelerating GEMM, and widely deployed in industry.
In this work, we describe two significant improvements to the traditional SA architecture, to specifically optimize for CNN inference.
Firstly, we generalize the traditional scalar PE, into a \textit{Tensor-PE}, which gives rise to a family of new \textit{Systolic Tensor Array} (STA) microarchitectures.
The STA family increases intra-PE operand reuse and datapath efficiency, resulting in circuit area and power dissipation reduction of as much as 2.08$\times$ and 1.36$\times$ respectively, compared to the conventional SA at iso-throughput with INT8 operands.
Secondly, we extend this design to support a novel block-sparse data format called density-bound block (DBB).
This variant (STA-DBB) achieves a 3.14$\times$ and 1.97$\times$ improvement over the SA baseline at iso-throughput in area and power respectively, when processing specially-trained DBB-sparse models, while remaining fully backwards compatible with dense models.

\end{abstract}

}

\maketitle

\IEEEdisplaynontitleabstractindextext

%
\IEEEpeerreviewmaketitle

\section{Introduction}
\label{sec:intro}

There is currently huge interest in accelerating Convolutional Neural Network (CNN) inference on mobile devices, for tasks such as image classification, detection and segmentation.
CNN layers are typically implemented by lowering 2D convolution to general matrix multiply (GEMM) kernels, which are typically the runtime bottleneck when executed on CPUs, motivating hardware acceleration.
The systolic array (SA) is a special-purpose processor for efficiently accelerating GEMM.
The SA consists of an array of MAC processing elements (PEs), which communicate operands and results using local register-to-register communication only, which makes the array very efficient and easily scalable without timing degradation.
These advantages have led to their deployment in commercial products, e.g. the Google Tensor Processing Unit (TPU)~\cite{tpu-isca-short}.


In this paper, we describe new SA microarchitectures specifically targeting mobile CNN inference with narrow INT8 operands, summarized below:
\begin{itemize}
\item{
The classical SA is generalized into a family of \textit{Systolic Tensor Array} (STA) by replacing the traditional scalar PE with \textit{Tensor-PEs}.
STAs have twice the area efficiency of the baseline SA at iso-throughput.
}
\item{
We introduce a novel \textit{Density-Bound Block} (DBB) structured-sparse matrix, which vastly improves load balancing, with negligible accuracy loss on five CNNs.
}
\item{
The STA is extended into STA-DBB, which supports DBB-sparse models.
Results demonstrate $3\times$ area efficiency and $\sim 2 \times$ power efficiency of the iso-throughput SA baseline. 
STA-DBB also outperforms SMT-SA~\cite{smt-sa_short} for INT8 operands, while remaining backward compatible with conventional (dense) CNNs.
}

\end{itemize}

\section{Background and Related Work}
\label{sec:background}

\begin{figure}[t]
\centering
\includegraphics[width=0.42\textwidth]{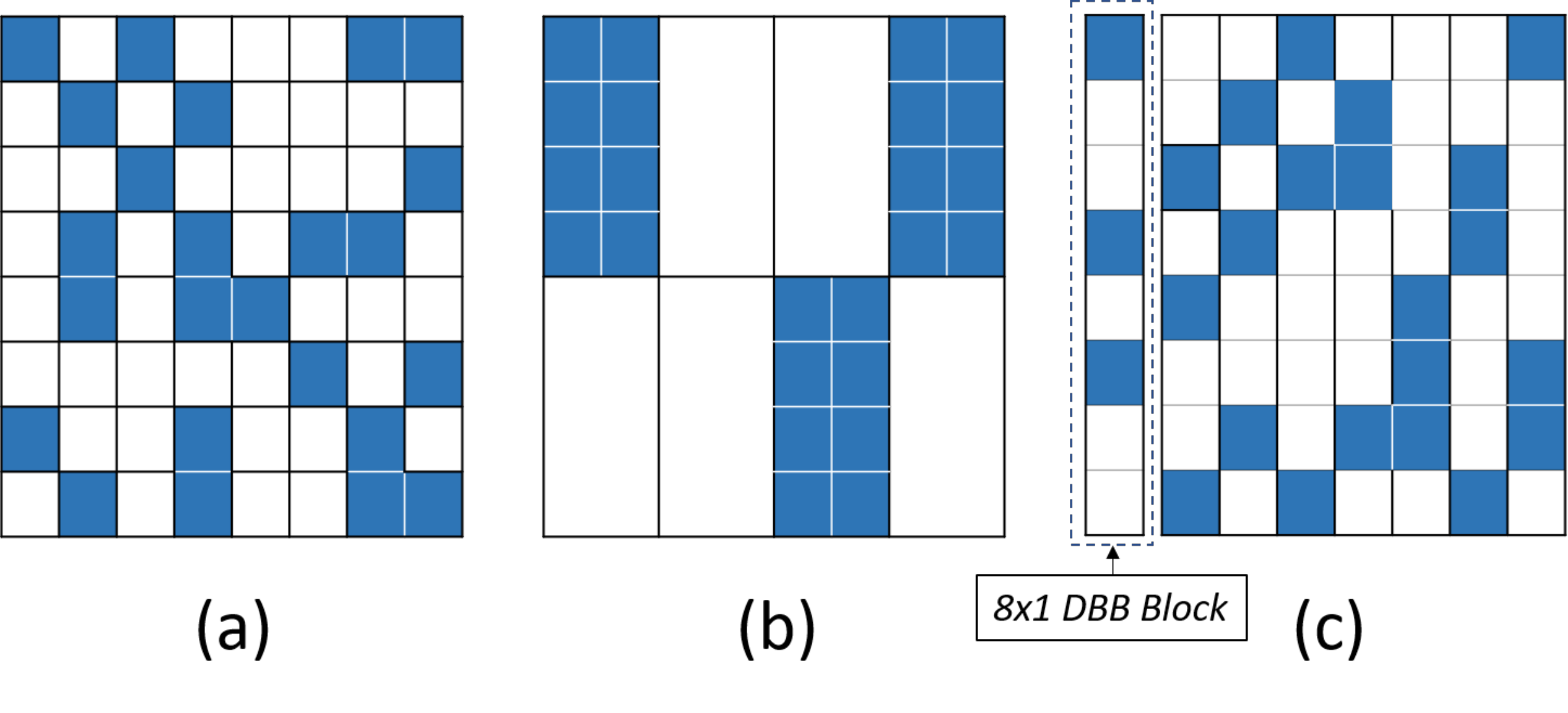}
\vspace{-10pt}
\caption{Sparse matrices: (a) random sparse; (b) block sparse (4$\times2$ block); (c) density-bound block (DBB) ($8\times1$ block). Filled elements denote non-zero. }
\label{fig:sparse_mats}
\end{figure}

\textbf{Systolic Arrays for CNN Inference.} 
Although the GEMM kernel is obviously relevant to a wide range of workloads, we focus here specifically on INT8 data type variants for CNN inference~\cite{tpu-isca-short}.
INT8 microarchitectures are especially challenging to optimize, as the datapath cost is relatively low compared to data movement, in contrast to floating-point data types which have a much higher relative datapath cost.
Our starting point is a TPU-like baseline~\cite{tpu-isca-short}, modified with: (1) clock gating on zero operands, and (2) output-stationary dataflow, which keeps the larger INT32 accumulators in place and instead shifts the smaller INT8 operands.

\textbf{Sparse Matrices for CNNs.}
CNN layers typically have sparsity in both the weight data (known at compile time) and activation data (known only at runtime).
These zero operands can be exploited by skipping operations with at least one zero.
This is a very compelling optimization because GEMM kernels are typically compute bound ($O(N^3)$).
Therefore, any computation on sparse data can potentially achieve an increase in throughput and/or power consumption.
Naturally occurring sparsity is often referred to as \textit{random sparsity} because there is no constraint on the locations of the zeros (Fig.~\ref{fig:sparse_mats}(a)).
Random sparsity can be very challenging to exploit, as (1) it requires indexes to be computed and communicated for each data item, and (2) load balancing challenges can lead to low utilization.
In contrast, structured \textit{block-sparsity}~\cite{matrix_pivoting_short,block_sparsity} (Fig.~\ref{fig:sparse_mats}(b)) groups zero weights into coarse-grained blocks.
Block sparse approaches are compelling, but impose a strict structure which results in poor accuracy on convolutional layers when the block size is sufficiently large to achieve an advantage in the hardware.
Building on previous work, this paper proposes the density-bound block (DBB) matrix format of Fig.~\ref{fig:sparse_mats}(c).

\textbf{Sparse Matrix Multiplication Accelerators}.
An effective approach to exploit zero operands in a \textit{scalar PE} is to clock-gate the input operand registers to reduce datapath toggling and therefore dynamic power (e.g. Eyeriss~\cite{eyeriss_short}).
An alternative is traditional sparse linear algebra, which involves storing and computing on explicit indexes that accompany the non-zero data.
For example, EIE~\cite{eie_short} implements a sparse matrix-vector  accelerator for fully-connected layers, and SCNN~\cite{scnn_short} implements indexed sparse CNN layers.
However, index-based approaches have significant overhead for storing and computing on the indexes themselves (e.g. a 10-bit index for each 16-bit element in EIE).
The overhead is especially severe for the common case for CNNs: ``medium'' sparsity INT8 weights.
In contrast, fixed CNN feature extractors (FixyNN~\cite{fixy2019sysml_short}) can very efficiently exploit random sparse weights.

Relating more specifically to systolic arrays, Kung et al.~\cite{kung_2018_short} demonstrated column combining of sparse matrices, before processing on an SA architecture.
While Shomrom et al.~\cite{smt-sa_short} describe a method to process multiple sparse matrices simultaneously in direct analogy to simultaneous multithreading (SMT) on CPUs.
In common with these papers, we build on the basic SA architecture, but instead of random sparsity, we exploit a novel structured-sparse approach with a Tensor-PE.

\section{Dense GEMM Accelerators}
\label{sec:hw}

\begin{figure*}[t]
    \centering
    \subfloat[Conventional Systolic Array (\textbf{SA})]{
        \centering
        \includegraphics[width=0.64\columnwidth]{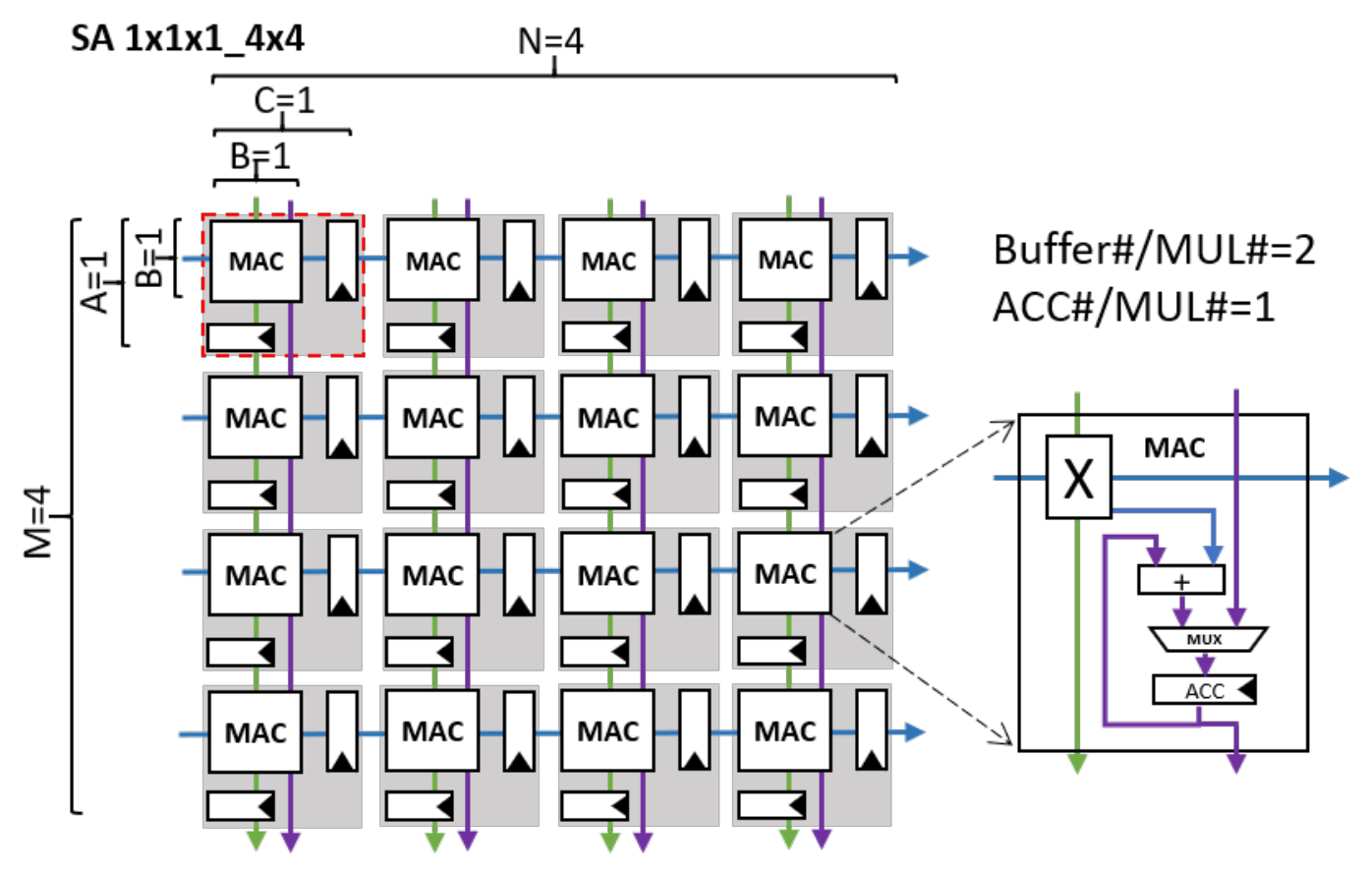}
        \label{fig:array:sa}
    }
    \hspace{1pt}
    \subfloat[Systolic Tensor Array (\textbf{STA})]{
        \centering
        \includegraphics[width=0.64\columnwidth]{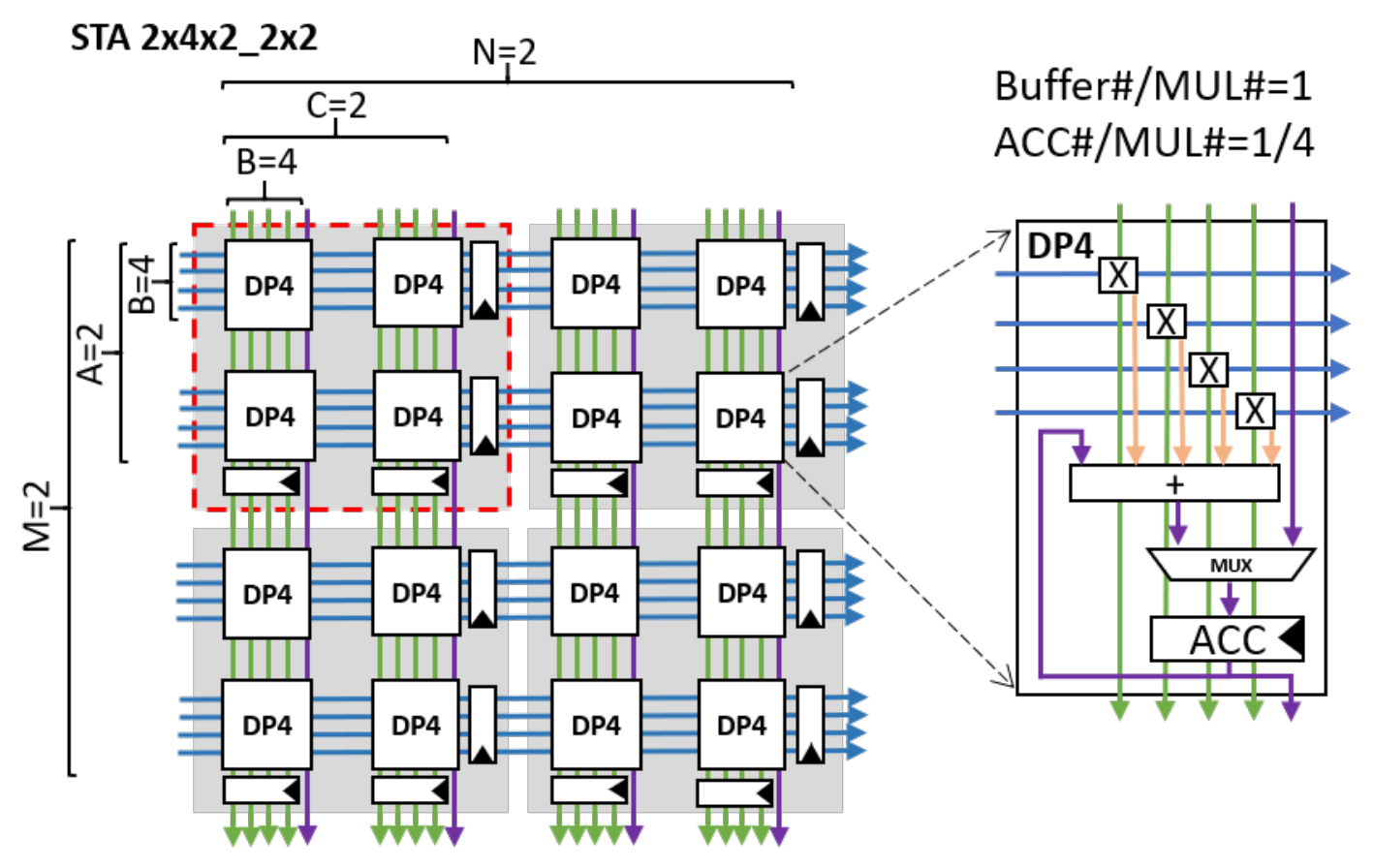}
        \label{fig:array:fmsa}
    }
    \hspace{1pt}
    \subfloat[Systolic Tensor Array for DBB (\textbf{STA-DBB})]{
        \centering
        \includegraphics[width=0.64\columnwidth]{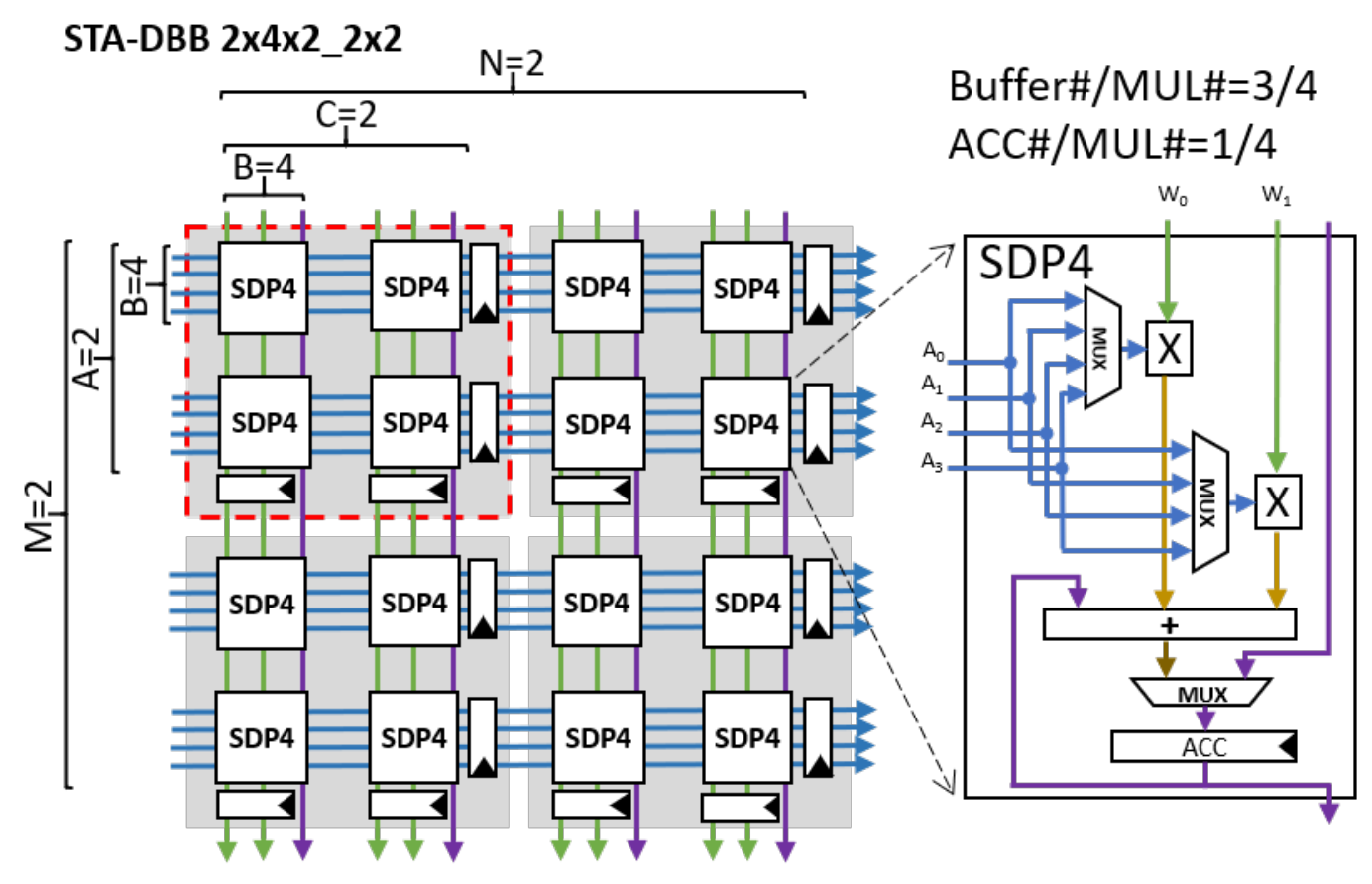}
        \label{fig:array:fmssa}
    }
    \caption{
    Systolic array (SA) microarchitectures, generalized by extending each scalar PE to a tensor PE which performs a matrix multiplication
    on each clock cycle. 
    Notation: A$\times$B$\times$C\_M$\times$N denotes a M$\times$N 2-D array of A$\times$B$\times$C tensor PE (dashed box). 
    Pipeline registers connect adjacent PEs, with only local data movement, which allows M$\times$N to scale without timing degradation.
    }
    \label{fig:array}
\end{figure*}

\subsection{Conventional Systolic Array (SA)}

Fig.~\ref{fig:array:sa} shows the classic SA, widely deployed in products.
We develop our architecture on top of the SA, and use it to baseline our results.
We target mobile vision applications, with INT8 operand registers ($REG$), and INT32 accumulator registers ($ACC$).
$M$ and $N$ describe the height and width of the PE array, respectively.
The top-to-bottom paths required to read out the spatially-distributed accumulators are not illustrated.

SAs are highly efficient due to very high operand reuse through local register-to-register communication.
This is in contrast to repeated SRAM reads for a dot-product machine~\cite{eie_short}, or global configurable communication for a generalized spatial array~\cite{eyeriss_short}. 
However, there is significant room for improvement in the SA architecture by relaxing the ratio of operand registers ($REG$) to MACs.
Each MAC operation traditionally requires two INT8 operand registers and one INT32 accumulator.

\subsection{Systolic Tensor Array (STA)}
\label{sec:hw:fm-sa}

We generalize the conventional SA (Fig.~\ref{fig:array:sa}) into the \textit{Systolic Tensor Array} (STA, Fig.~\ref{fig:array:fmsa}), by fusing a block of scalar PEs into a single \textit{tensor PE} .
Each STA architecture contains 
$M \times N$ tensor PEs, with each tensor PE consisting of a sub-array of $A \times C$ MACs that each perform a dot-product operation on $B$ operand pairs.
This is denoted uniquely as A$\times$B$\times$C\_M$\times$N.
Fig.~\ref{fig:array:fmsa} shows a $2\times 2$ array of tensor PEs, each with a $2\times 2$ datapath of 4 operand pair dot-product units ($DP4$).
Note that the classic SA (Fig.~\ref{fig:array:sa}) is a special case of STA, with $A=B=C=1$ (denoted 1$\times$1$\times$1\_M$\times$N).

Fig. \ref{fig:dataflow} shows the data flow of a $4 \times 4$ by $4\times4$ matrix multiplication on  a 2$\times$2$\times$2\_2$\times$2 STA. 
A significant efficiency improvement is achieved here by reducing the number of operand buffers per MAC by $2 \times$, and the number of accumulator buffers per MAC by $4\times$.
To our knowledge, tensor PEs have not been previously described in the SA context.



\section{Sparse GEMM Accelerators for CNNs}
\label{sec:sparse}
As described in Section~\ref{sec:background}, both random and block sparse approaches introduce significant trade-offs -- random sparsity is hard to exploit in hardware, and block sparsity significantly impacts accuracy in CNNs.
In this section, we propose an alternative to the conventional block sparse matrix format.

\subsection{Density-Bound Block (DBB)}
\label{sec:dbb}


\begin{figure}[t]
\centering
\includegraphics[width=0.47\textwidth]{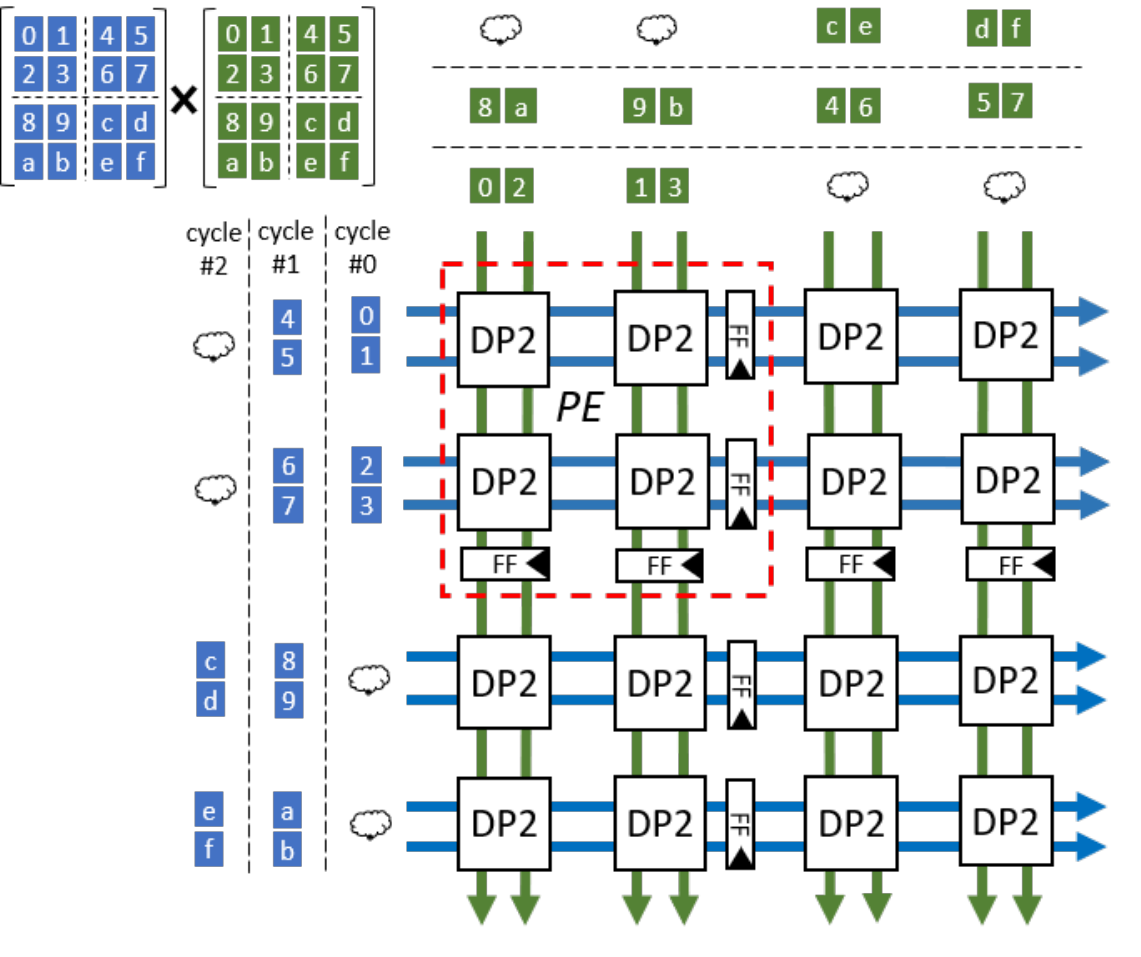}
\vspace{-8pt}
\caption{
Example 2$\times$2$\times$2\_2$\times$2 STA data flow. 
The tensor PE (red dashed box) computes a 4$\times$4 by 4$\times$4 matrix multiplication.
Each input operand matrix is split into a 2$\times$2 sub-matrix, 
with columns (rows) of data in blue (green) fed into the array from the left (top) edges over each clock cycle.}
\label{fig:dataflow}
\end{figure}

Fig.~\ref{fig:sparse_mats}(c) shows the proposed density-bound block (DBB) matrix, which simply places an upper limit on the number of non-zero ($NNZ$) elements in each block. 
For example, Fig.~\ref{fig:sparse_mats}(c) consists of 8 blocks of 8$\times$1, each with up to 3 non-zero values ($NNZ$ $\le$ 3). 
This is in contrast to conventional sparse block (Fig.~\ref{fig:sparse_mats}(b)), where each block is either entirely unconstrained or all zero. 
DBB is a middle-ground between random (Fig.~\ref{fig:sparse_mats}(a)) and block (Fig.~\ref{fig:sparse_mats}(b)) sparse formats, and results in higher CNN accuracy with the same $NNZ$, because the distribution of non-zero elements is significantly less constrained.
At the same time, the compute required from the hardware is known a-priori for each block and high utilization is guaranteed.
A simple bitmask compression is used to encode each block of 8 elements (one byte overhead per block), along with the four bytes of non-zero data.
This yields a 37.5\% reduction in weight memory footprint.

\subsection{Systolic Tensor Array for DBB (STA-DBB)}
\label{sec:hw-sparse}

Next, we add support to the STA architecture for DBB-sparse weight matrices.
Weight sparsity is easier to exploit than activations, as the values are known ahead of time.
The DBB weight matrix upper bounds the number of non-zero operands.
Therefore, the number of physical MAC units required is no greater than the DBB sparsity bound.
For instance, if the DBB block size is 8 with $NNZ$ $\le$ 4, each 8-input Dot Product unit 
($DP8$) only requires 4 MAC units instead of 8, which represents a 50\% reduction in MAC hardware at the same throughput. 
This approach requires a multiplexer in font of MAC to select the corresponding input feature map (activation) based on the index of the non-zero weight. 

Fig.~\ref{fig:array:fmssa} illustrates a 2$\times$4$\times$2\_2$\times$2 Sparse STA for DBB (STA-DBB) with 4-input Sparse Dot Product ($SDP4$) unit, which has 4-value vector activation inputs $[A_0,A_1,A_2,A_3]$ from left and a 50\% DBB compressed weight inputs $[W_0, W_1]$ and associate 2-bit non-zero indices from top and accumulates the dot product into accumulator ($ACC$). 
In each SDP4, we trade two 8-bit multipliers for two 8-bit 4:1 multiplexers (MUX), where a MUX costs significantly less than an 8-bit multiplier.
The array performs 16 effective MACs per clock cycle with only 8 physical multipliers. 
The final results in the stationary accumulators are read out from bottom of the array through shift chains in four clock cycles, after the computation is finished.
The data flow is similar to Fig.\ref{fig:dataflow}, except that the matrix input from the top is $50\%$ DBB-sparse.
Notably, this architecture still supports conventional dense GEMM at half throughput, which will probably remain important to support the widest range of workloads.


\section{Evaluation Results}
\label{sec:results}

\subsection{DBB-Sparse CNN Training Results}

\begin{table}
\caption{CNNs trained with 8-bit DBB-sparse weights.}
\centering
\begin{tabular}{l c c c c}
\toprule
Model                                & Dataset           & Baseline    &  \multicolumn{2}{c}{Pruned Model Result}  \\
                                     &                   & Acc.(\%)    & Acc.(\%)  &  NNZ (\%)$^1$         \\
\midrule                            
LeNet-5 (DBB)                        & MNIST             & 99.1        &  98.7      & 1.05K (25)            \\ 
ConvNet (DBB)                        & CIFAR10           & 86.0        &  85.3      & 26.8K (25)            \\
MobileNetV1 (DBB)                    & ImageNet          & 70.9        &  69.8      & 1.6M (50)             \\
ResNet-50V1 (DBB)                    & ImageNet          & 75.2        &  74.2      & 8.79M (37.5)          \\
VGG-16 (DBB)                         & ImageNet          & 71.5        &  71.4      & 5.39M (37.5)          \\
\midrule                            
AlexNet \cite{SSL-alexnet-short}           & ImageNet          &  57.4       &  57.5      & 2.81M (75)            \\ 
VGG-19$^2$ \cite{kung_2018_short}    & ImageNet          &  --$^3$       &  71.8      & 4.1M (12.6)           \\
\bottomrule
\end{tabular}
\vspace{2pt}

$^1$Convolution layers.
$^2$Modified VGG-19, $\sim$31M conv param. 
$^3$Not reported.

\label{tab:dbb-training}
\end{table}

\iftrue
\begin{figure*}[ht]
\centering
\includegraphics[width=1.0\textwidth]{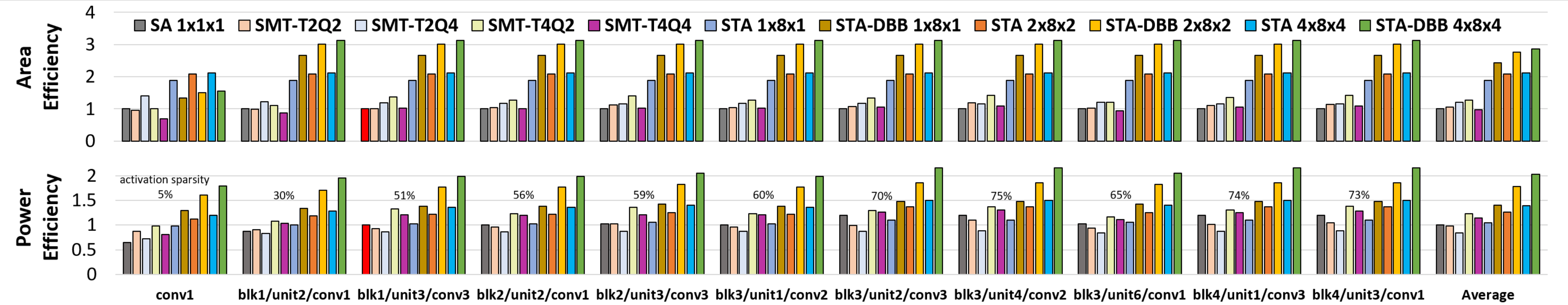}
\caption{Normalized area and power efficiency (iso-throughput, higher is better) for typical layers of ResNet50\_V1 model, trained at 8-bits, 62.5\% sparse weight, 1$\times$8 DBB (conv1 remains dense). Area and power efficiency normalized to SA 1x1x1 with 50\% average activation sparsity, closest to the  $\textit{blk1/unit3/conv3}$ layer in this ResNet50 example.
}
\label{fig:resnet_v1_layers}
\end{figure*}
\fi

To demonstrate the feasibility of DBB models, we trained five CNNs, applying  conventional INT8 quantization and amplitude-based pruning for VGG-16, MobileNetV1, ResNet-50V1, 5-layer ConvNet and LeNet-5 on ImageNet, CIFAR10 and MNIST datasets.  
The validation results  are given in Table~\ref{tab:dbb-training}. 
The accuracy loss is in the range of 0.1\% to 1.1\% across all five DBB models, which include both relatively big examples (ResNet-50V1) and parameter-efficient models (MobileNetV1).


\iftrue

\begin{figure}[ht]
    \centering
        \includegraphics[width=1.0\columnwidth]{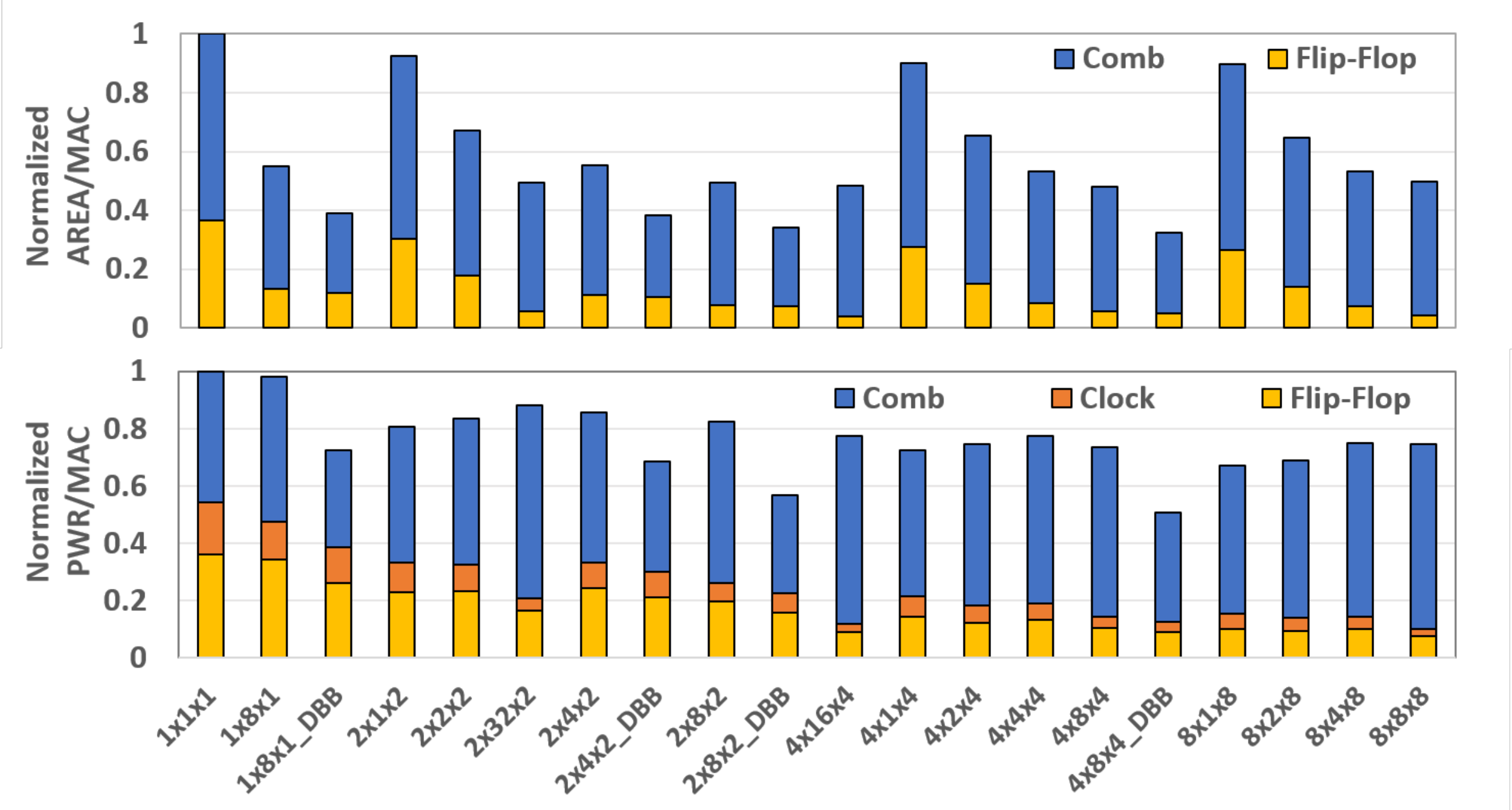}
    \caption{Area and power at iso-throughput (lower is better) for STA and STA-DBB (50\% sparse), with cell breakdown.} 
    \label{fig:ppa}
\end{figure}
\fi

\subsection{Hardware Accelerator Results}
\label{sec:results:hw}

In this section, we evaluate four architectures: (1) conventional TPU-like SA baseline (1$\times$1$\times$1) with a dense model; (2) optimized STA with dense model; (3) STA-DBB with a 50\% DBB-sparse model, and (4) SMT-SA with 62.5\%-sparse model~\cite{smt-sa_short}.
Each design was implemented in Verilog using a parameterized RTL generator, and synthesized in a TSMC 16nm FinFET process using Synopsys Design Compiler with a 1GHz clock. 
For power simulation we use Synopsys PrimeTimePX with estimated clock tree and fully-annotated switching activity.

\begin{table}[t]
\caption{Throughput-normalized area and power efficiency with 50\% sparse activation at 1GHz. Normalized to SA 1x1x1 (baseline).}
\vspace{-8pt}
\begin{center}
\begin{tabular}{l c c c c}
\toprule
Design & Model Sparsity & Array & Area Eff.$^1$ & Power Eff.$^1$ \\
\midrule
SA-NCG$^2$~\cite{tpu-isca-short}    & Dense           & 1$\times$1$\times$1     & 0.95    & 0.65  \\
SA$^3$                      & Dense           & 1$\times$1$\times$1     & 1.00    & 1.00  \\
STA                       & Dense           & 4$\times$8$\times$4     & 2.08    & 1.36 \\
SMT-SA~\cite{smt-sa_short}  & Random (62.5\%) & T2Q4                    & 1.21    & 0.80  \\
STA-DBB                      & DBB (50\%)      & 4$\times$8$\times$4     & \textcolor{red}{3.14} & \textcolor{red}{1.97} \\
\bottomrule
\end{tabular}
\end{center}
$^1$ Throughput normalized.
$^2$ SA no clock gating.
$^3$ Baseline clock-gated SA.
\label{tab:ppa}
\end{table}

Fig.~\ref{fig:ppa} shows area and power results for the baseline dense SA (with and without clock gating for zero operands), and improved STA designs, across the design space of tensor-PE dimensions that meet a 1GHz clock frequency.
Results confirm that the traditional SA (1$\times$1$\times$1) has 36\% area and 54.3\% power attributed to registers alone.
The optimized STA designs show improvements of as much as 47\% area and 73\% power at the design space sweet spot (4$\times$8$\times$4).
The flip-flop reduction shown by the blue portions in Fig.~\ref{fig:ppa}, comes from reduced operand pipeline buffers and accumulator flip-flops sharing. 
The reduction in combinational logic (yellow), is due to efficiencies of the adder tree in the tensor MAC (dot-product).

STA-DBB further enhances efficiency by taking advantage of DBB-sparse models.
Fig.~\ref{fig:ppa} shows the STA-DBB designs with DBB-sparse weight compression reduce the area and power per MAC by up to 30\%, over and above corresponding FM-SA configuration. 
The reduction in combinational logic (blue), comes from the 50\% reduction in physical multipliers, while the reduction in registers (yellow) and clock load (orange), is from pipeline buffers saving. 

We also evaluated Simultaneous Multi-Threading Systolic Arrays (SMT-SA)~\cite{smt-sa_short}, which process random-sparse weights, and use a FIFO to overcome the load imbalance challenge.
We implemented the T2Q2 configuration, which denotes 2 threads with a 2-deep FIFO queue, as well as the
T2Q4, T4Q2 and T4Q4, all with INT8 operands and INT32 accumulation.
Note that for INT8, SMT-SA which exploits random sparsity is actually less efficient than STA, which doesn't even exploit sparsity.
This is due to the overhead that the FIFOs introduce, relative to the size of the INT8 datapath logic.

Fig.~\ref{fig:resnet_v1_layers} shows the efficiency of some typical layers from the ResNet50\_v1 model. 
SMT-SA area and power efficiencies are $0.8\times$-$1.21 \times $ the SA baseline and around 2$\times$ lower than the corresponding STA and STA-DBB architectures with 62.5\% sparse weights and 39-75\% sparse input feature maps. 
The high hardware cost of the FIFO in SMT-SA designs cancels out the benefit of sparsity for INT8 operands.   

Table~\ref{tab:ppa} summarizes the area and power efficiency results of the best configurations of the four architectures studied.

\section{Conclusion}
\label{sec:conclusion}

The systolic array (SA) is an efficient, highly parallel architecture widely used to compute matrix multiply. 
In this paper, we improved on the classic SA architecture by specifically optimizing for efficient CNN inference. 
Firstly, we generalize the traditional SA by introducing a tensor PE with multiple parallel operations.  
This systolic tensor array (STA) architecture significantly reduces circuit area and power dissipation by as much as 2.08$\times$ and 1.36$\times$ respectively compared a conventional clock gated SA, due an increase in intra-PE operand reuse and datapath efficiency, reducing operand register count and clock tree load. 
Secondly, we further extend the STA architecture to efficiently accelerate models trained with density-bound block (DBB). 
This new sparse architecture (STA-DBB) shows 3.14$\times$ and 1.97$\times$ improvement in area and power respectively when processing specially trained DBB-sparse models, while remaining fully backward compatible with traditional dense models.


%

%
%

%
%
%
%




\bibliographystyle{IEEEtran}


\bibliography{dnn}

\end{document}